\documentclass[final,5p,times,twocolumn]{elsarticle}

\usepackage{amssymb}
\journal{Journal of Quantitative Spectroscopy and Radiative Transfer}

\begin{document}

\begin{frontmatter}



\title{Terahertz binding of nanoparticles based on graphene surface plasmons excitations}

%%\author[label1,label2]{}
\author[label1,label2]{Hernan Ferrari }
\address[label1]{Consejo Nacional de Investigaciones Cient\'ificas y T\'ecnicas (CONICET).}
\address[label2]{Facultad de Ingenier\'ia, Universidad Austral, Mariano Acosta 1611, Pilar 1629, Buenos Aires, Argentina.}
\author[label3]{Carlos J. Zapata--Rodr\'guez}
\address[label3]{Department of Optics and Optometry and Vision Sciences, University of Valencia, Dr. Moliner 50, 46100, Burjassot, Spain.}
\author[label1,label2]{Mauro Cuevas \corref{cor1}}
\ead{mcuevas@austral.edu.ar}

%\address[label4]{Consejo Nacional de Investigaciones Cient\'ificas y T\'ecnicas (CONICET)}

\cortext[cor1]{corresponding author}
\author{}

\address{}

\begin{abstract}
This work studies the optical binding of a dimer composed by  %identical 
dielectric particles close to a graphene sheet.  Using a rigorous electromagnetic method, we calculated the optical force acting on each nanoparticle. In addition, we deduced analytical expressions enabling to evaluate the contribution of graphene surface plasmons (GSPs) to optical binding. Our results show that surface plasmon on graphene excitations generate multiple equilibrium positions for which the distance between particles are tens of times smaller than the photon wavelength. Moreover, these positions can be dynamically controlled by adjusting the chemical potential on graphene. Normal and oblique incidence have been considered.
\end{abstract}


\end{frontmatter}


%\section{Introduction} \label{introduccion}

\section{Introduction} 

The scattered field by a neutral particle assembly  can result, under particular conditions, in an effective dipole--dipole interaction useful for optically binding the particle array. Most of the literature in this field has been devoted to micro--sized particles,  since the significant electromagnetic scattering at this scale makes the strength of optical binding surpassing the hydrodynamic interactions and the stochastic thermal activation \cite{Dholakia,Forbes}. 

The significant progress made in experimental techniques and the extensive wealth of theoretical research have meet the need of optical binding between nano sized particles.  In this context, the effect of localized surface plasmon excitations on assembly clusters of metallic nanoparticles has been theoretically and experimentally demonstrated \cite{NV1,Yan,Salary,Demergis}. In addition, stable optical binding between dielectric nano particles using an evanescent field formed by total reflection at a dielectric interface has been demonstrated \cite{EV1,EV2}.  
Recently, a planar metallic or metamaterial structure was proposed as an optical binding tool, in which the propagating eigenmodes excited on the surface play an essential role on the increment of the interaction force between dielectric nano particles. The binding distance between nanoparticles is defined by the eigenmode wavelength which, in case of bound modes such as surface plasmons, is smaller than the photon wavelength \cite{NV2,kostina}. 

It is known, apart from the well known surface plasmons supported by a metallic surface, long livid propagating plasmons can be supported by graphene from terahertz (THz) to infrared (IR) frequency range. % Some outstanding properties of 
GSPs present good tunability through electrical or chemical modification of the carrier densitiy, relative low loss and tightly confined fields. %, relative low loss, and good tunability from terahertz up to mid--infrared frequencies  through electrical or chemical modification of the carrier densitiy [citas]. 
These properties have been the subject of theoretical and experimental studies to  find application in a wide range of disciplines. For example, GSPs have been proposed as a new generation of nano  antennas for communications in the THz bands \cite{jornet}, molecular sensors capable to selectively enhance individual spectral features \cite{filter,nong},  low frequency spaser (see \cite{leila} and Refs. therein) and for enhancing the spontaneous emission and energy transfer between molecules \cite{Debu,OC,Bradshaw}. 
Recently, the use of high-quality GSP resonances on twin graphene--coated dielectric rods to enhance the diffraction radiation intensity in the IR  region has been demonstrated \cite{nosich}. 

Regarding to optical tweezers framework, some works have focused on the optical trapping using graphene as plasmonic material  \cite{OT1menos,OT0}. In these structures, GSPs provided the channels for capturing a small
dielectric nanoparticle \cite{OT1} or trapping and sorting nanoparticles \cite{OT2}.    
%
%In \cite{OT1} it has proposed a periodic graphene nanogap array for  capturing a small dielectric nanoparticle and, in \cite{OT2} a structure formed by two parallel graphene stripes has been disigned as an efectively trapping and sorting nanoparticles. 
%
In \cite{OT22}, authors proposed a nanopatterned graphene structure for particle
trapping/levitation.  In addition, optical force on the nonlinear graphene--wrapped nanoparticle has been investigated \cite{OT3}.      

%In this work we propose a plasmonic nanotweezers that works in the THz range to overcome the above-mentioned challenges.

%In this context, 
A natural question that arises from the above outstanding characteristics is whether propagating GSPs  can be useful for binding particles, taking advantage of their great localization to bind particles greatly exceeding the diffraction limit. In particular, how GSPs enable dielectric nanoparticles to be bound %how possible could be to bind dielectric nanoparticles 
at distances of few hundred nanometers between them by using low frequency radiation instead of visible radiation. %In order to respond these questions, the present work is theoretically developed covering . 
In consequence, the present work is theoretically developed in order to respond to these questions. 
%and the growing demands about . 
%
%In consequence, the development of efficient nitrate sensors appears as a necessity in order to respond to the growing demand on the better … of AgNPs obtained with 0.5 molar equivalent of oleic acid This color is characteristic of the localized surface plasmon resonance (LSPR %
%
Specifically, we analytically study the binding conditions for a dimer formed by dielectric nanoparticles,   providing a complete description about the use of GSPs to create equilibrium positions solely by plane wave radiation. Since the kinematic properties of GSPs can be controlled by chemical potential variations,  possibilities to dynamically control the equilibrium   positions without changing the geometrical parameters of the structure are opened. 

This paper is organized as follows. In section 2, we present a brief description of 
the analytical method, based on the Green approach,  to calculate the optical force between two dielectric nanoparticles placed above a graphene sheet.  By using contour integration in the complex plane,
we obtained simple formulas, which contain the GSP contribution to the optical force, that reproduce the main results obtained by applying the rigorous theory.  In section 3,  we %study the cases in which the  field is normally and obliquely  incident  on  the graphene plane 
exemplify 
 for both normally and obliquely incident
illumination cases, and for different particle shapes. %  we present numerical results calculated under two  orthogonal incident electric field orientations. 
Concluding
remarks are provided in section 4. The international system of units is used and an $\exp(-i \omega t)$ time--dependence is implicit
throughout the paper, where $\omega$ is the angular frequency, $t$ is the time coordinate, and $i =\sqrt{-1}$. The symbols Re and Im are used for denoting the real and imaginary parts of a complex
quantity, respectively.

%Decir que se obtienen formulas analiticas que reproducen los principales resultados obtenidos numericamente.

%Decir que se obtienen formulas analiticas que reproducen los principales resultados obtenidos numericamente.

\section{Theory}\label{teoria}

\subsection{Optical binding.}

We consider a system formed by two dielectric % and spherical 
particles placed above of a plane graphene sheet (at $z=0$) separating two homogeneous half spaces, vacuum ($\varepsilon_1=1$) and a non--magnetic and isotropic dielectric medium ($\varepsilon_2=2.13$).  Both dielectric particles are identical. Spherical particles are characterized by radius  $R_1=R_2=R$, and other shapes  are taken in such a way the volumes of these objects are equal to that of the sphere. The relative permittivity of the particles is  $\varepsilon_p$ and they are placed at $\mathbf{r}_A=z_0 \hat{z}$   and $\mathbf{r}_B=  x_0  \hat{x}+z_0 \hat{z}$ ($z_0 \ge 0$) in the field of an incident plane wave (see Fig. 1). 
\begin{figure}[htbp!]
    \centering
%    \subfloat[]
    {\includegraphics[width=0.35\textwidth]{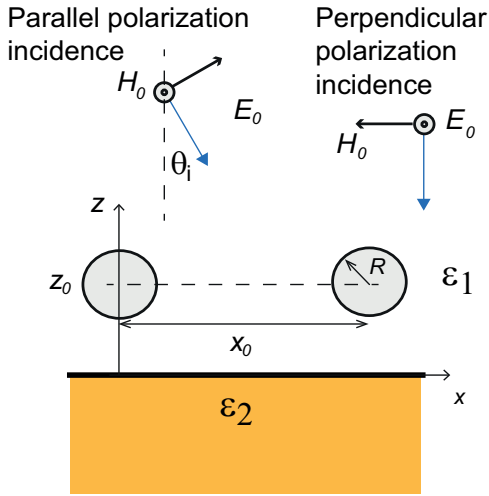}}
%    \hspace{1mm}
%    \subfloat[]{\includegraphics[width=0.45\textwidth]{OSA2.png}}
    \caption{Schematic of the problem. Two identical particles of permittivity $\varepsilon_p$  under illumination of a plane wave. The particles are embedded in vacuum ($\varepsilon_1=1$) at a distance $z_0$ from the graphene plane, at $z=0$, that separates the vacuum from a dielectric medium characterized by a permittivity  $\varepsilon_2=2.13$. In case of normal incidence, the electric field can be parallel (parallel polarization) or perpendicular (perpendicular polarization) to the axis joining the particles.  }
    \label{sistema}
\end{figure}

Taking into account that the size of the particles are smaller than plasmon and photon  wavelengths  ($R<\lambda_{sp}<<\lambda$), we can use the dipolar approximation. In this framework,  the time average of the total force acting on a single particle is \cite{NV},
\begin{equation}\label{eq1}
\mathbf{F(\mathbf{r})} = \frac{1}{2} \mbox{Re} \sum_{j=x,y,z} p_j^* \nabla E_j(\mathbf{r}), 
\end{equation}
where $E_j$ is the $j$ component of the electric field, $\mathbf{p}=\alpha_0 \mathbf{E}$ is the induced electric dipole on particle at $\mathbf{r}$ position, 
\begin{equation}\label{alpha_0}
\alpha_0=\frac{\alpha_e}{1-i\frac{k_0^3}{6 \pi \varepsilon_0} \alpha_e},
\end{equation}
is the radiation corrected electric polarizability, $k_0=2\pi/\lambda$ is the modulus of the  vacuum photon  wavevector, $\varepsilon_0$ is the vacuum permittivity and $\alpha_e$ 
is the electrostatic approximation of particle polarizability. In case of spherical particle, 
\begin{equation}\label{alphae}
\alpha_e=4\pi \varepsilon_0 R^3 \frac{\varepsilon_p-\varepsilon_1}{\varepsilon_p+2\varepsilon_1}.
\end{equation}
In case of other shapes such as cubes or cylinders particles, we have calculated the $\alpha_e$ polarizability following the accurate formulas presented in   \cite{cubo_cilindro}.  

The electric field in Eq. (\ref{eq1}) is given by
\begin{equation}\label{eq2}
\mathbf{E(\mathbf{r})} = \mathbf{E}_0 + \frac{k_0^2}{\varepsilon_0} \mathbf{\hat{G}}(\mathbf{r},\mathbf{r}_A) \mathbf{p}_A + \frac{k_0^2}{\varepsilon_0} \mathbf{\hat{G}}(\mathbf{r},\mathbf{r}_B) \mathbf{p}_B, 
\end{equation}
where $\mathbf{E}_0$ is the superposition of the incident field and the reflected field on the flat substrate, %is the sum of the incident and reflected by the surface fields,
$\mathbf{\hat{G}}(\mathbf{r},\mathbf{r}_j)=\mathbf{\hat{G}}_0(\mathbf{r},\mathbf{r}_j)+\mathbf{\hat{G}}_s(\mathbf{r},\mathbf{r}_j)$ is the sum of the vacuum and the  scattered--by--the--surface   Green tensors of a point dipole. Second and third terms in Eq. (\ref{eq2}) correspond to the scattered electric fields on $\mathbf{r}$ position by the electric dipole in particle A  and B, respectively. By solving the self consistent Eq. (\ref{eq2}) for $\mathbf{E}(\mathbf{r}_j)$ ($j=\mbox{A, B}$) and using Eq. (\ref{eq1}) we can calculate the optical force on particle A or B. An equivalent form using the effective particle polarizabilities defined in  \ref{Induced polarizability} that simplify the force calculation was considered in \cite{NV1,NV2}. For example, by using the expression for $\mathbf{p}_A$ given by Eq. (\ref{pA2}) and that equivalent for $\mathbf{p}_B$, Eqs. (\ref{eq2}) and Eq. (\ref{eq1}), we obtain an expression for the force on particle B,
\begin{eqnarray}\label{eq1B}
\mathbf{F(\mathbf{r}_B)} = \frac{1}{2} \mbox{Re} \sum_{j=x,y,z} p_{B j}^* \nabla E_j(\mathbf{r})|_{\mathbf{r}_B}=
\frac{1}{2} \mbox{Re}
\sum_{j=x,y,z} p_{B j}^* \nabla \Bigg\{ 
[\mathbf{E}_0(\mathbf{r})]_j \nonumber\\
+\frac{k_0^2}{\varepsilon_0} [\mathbf{\hat{G}}(\mathbf{r},\mathbf{r}_A)\mathbf{p}_A]_j + \frac{k_0^2}{\varepsilon_0} [\mathbf{\hat{G}}(\mathbf{r},\mathbf{r}_B) \mathbf{p}_B]_j\Bigg\}|_{\mathbf{r}_B}, 
\end{eqnarray}
where $[ \cdot ]_j$ indicates the $j$ coordinate of $\cdot$. An equivalent equation can be obtained for particle A. 

\subsection{Graphene surface plasmon contribution.}

Graphene surface plasmons, eigenmodes of the structure that propagate along the graphene surface  with their electric and magnetic fields decaying exponentially
away from the graphene sheet, can provide 
channels for enhancing the optical force between particles. 
Since Im $\sigma>0$ ($\sigma$ is the graphene conductivity) in the frequency range considered in this work, only $p$ polarized GSPs exist, \textit{i.e.}, surface waves with the total magnetic field parallel to the surface  \cite{milkhailov}.
 Thus, the full characteristics of the GSPs can be obtained by studying the singularities of the analytic continuation of the reflection coefficient $r_p(k_{||})$. % in Eq. (\ref{fresnelp}). 
 Pole singularities occur at complex locations 
and they represent the propagation  constant $k_{sp}$ of the GSPs. Since the integration path in Green's functions %(\ref{G3}) 
\begin{eqnarray} \label{G3p}
 \mathbf{\hat{G}}_s(\mathbf{r}_B,\mathbf{r}_A)=\frac{i}{8 \pi^2} 
 \int_0^\infty dk_{||} \, \mathbf{f}(k_{||},\rho)\,e^{i  \gamma^{(1)}(z_A+z_B)},
\end{eqnarray}  
is set along the
real and positive $k_{||}$ axis, the integral, and consequently the Green's functions, will be strongly
affected by singularities $k_{||}=k_{sp}$ that are close to that axis. In Eq. (\ref{G3p}),  $k_{||}$ is the wave vector parallel to the surface,   $\rho=\sqrt{|\mathbf{r}_A-\mathbf{r}_B|^2-(z_A-z_B)^2}$ and functions $\mathbf{f}(k_{||},\rho)$ are defined in \ref{Green}.  
 Following the same steps as in Ref. \cite{OZC}, we can  extract the GSP  contribution from the Green functions $\mathbf{G_s(\mathbf{r}, \mathbf{r}_j)}$ ($j=\mbox{A, B}$) as follows. % (see appendix \ref{GSP contribution}). 
Since the procedure is similar for all components of  $\mathbf{\hat{G}}_{s}(\mathbf{r},\mathbf{r}')$, we show it for the $xx$ component. Applying the symmetry properties of Bessel integrals \cite{OZC}, 
\begin{eqnarray}\label{paridad}
\int_0^{+\infty} f_{odd}(x) J_0(x) dx=\frac{1}{2}\int_{-\infty}^{+\infty} f_{odd}(x) H_0^{(1)}(x) dx ,\nonumber \\
\int_0^{+\infty} f_{even}(x) J_1(x) dx=\frac{1}{2}\int_{-\infty}^{+\infty} f_{even}(x) H_1^{(1)}(x) dx ,
\end{eqnarray}
where $f_{odd}(x)$ and $f_{even}(x)$ are odd and even functions, respectively, of the argument, we obtain
\begin{eqnarray} \label{Gxxsp}
 \hat{G}_{s,xx}(\mathbf{r}_B,\mathbf{r}_A)=\frac{i}{8 \pi^2} 
 \int_{-\infty}^\infty dk_{||} \, 2\pi \nonumber\\
 \times \frac{k_{||} \gamma^{(1)}}{k_0^2}  \frac{1}{2}[ \frac{H_1(k_{||} x_0)}{k_{||}x_0}-H_0(k_{||}x_0) ] r_p(k_{||})  \,e^{i 2 \gamma^{(1)}z_0}=\nonumber\\
 -\frac{i}{8 \pi^2} 
 \int_{-\infty}^\infty dk_{||} \, 2\pi 
 \frac{k_{||} \gamma^{(1)}}{k_0^2}  \frac{1}{2} H'_1(k_{||}x_0) r_p(k_{||})  \,e^{i 2 \gamma^{(1)}z_0},
\end{eqnarray}  
where we only have considered the $p$ polarization. In the last equality in Eq. (\ref{Gxxsp}) we used the fact that $H_1(z)/z-H_0(z)=-H'_1(z)$.  % (only $p$ polarized  GSPs exist for frequencies  below the graphene chemical potential). 
Note that  we have set $\theta=0$,  $z_A=z_B=z_0$ and $\rho=x_0$  to write Eq. ({\ref{Gxxsp}}) because both particles are on $x$ axis and at the same distance from the graphene plane. We now deform the integration path in (\ref{Gxxsp}) into a semicircle of large radius ($|k_{||}|\rightarrow \infty$) in the positive imaginary half--plane $\mbox{Im} k_{||}>0$, avoiding the branch point and pole singularities, as in Ref. \cite{OZC}. Then, the residues theorem gives
\begin{eqnarray} \label{Gxxsp2}
 \hat{G}_{s,xx}(\mathbf{r}_B,\mathbf{r}_A)=%- \frac{ k_{sp} \gamma^{(1)}_{sp}}{4 k_0^2} [ \frac{H_1(k_{sp}\rho)}{k_{sp}\rho}-H_0(k_{sp}\rho) ]   \,e^{i 2 \gamma^{(1)}z'}  \mbox{Res}\,r_p=\nonumber\\
   \frac{ k_{sp} \gamma^{(1)}_{sp}}{4 k_0^2}  H'_1(k_{sp}x_0)   \,e^{i 2 \gamma^{(1)}z_0}  \mbox{Res}\,r_p,
\end{eqnarray}  
where $k_{sp}$ is the propagation constant of GSPs,  $\gamma^{(1)}=\sqrt{k_0^2-k_{sp}^2}$ and 
\begin{eqnarray}\label{Res}
\mbox{Res}\, r_p= \lim_{k_{||}\to k_{sp}}(k_{||}-k_{sp}) r_p.
\end{eqnarray}
As $k_{sp}$ is almost real and higher than the modulus of the photon wave vector $k_0$, $\gamma^{(1)} \approx i \gamma_{sp}$ with $\gamma_{sp}=\sqrt{k_{sp}^2-k_0^2}$ a real number ($k_{sp}>k_0$). Thus, Eq. (\ref{Gxxsp2})  can be written as,
\begin{eqnarray} \label{Gxxsp3}
 \hat{G}_{s,xx}(\mathbf{r}_B,\mathbf{r}_A)=i 
  \frac{ k_{sp} \gamma_{sp}}{4 k_0^2} H_1'(k_{sp}x_0)   \,e^{- 2 \gamma_{sp}z_0}  \mbox{Res}\,r_p.
\end{eqnarray}  
Using the non retarded approximation, \textit{i.e.},  $k_0<<k_{sp}$, % in such away $k_0$ can be neglected, 
it follows     $\gamma_{sp}=k_{sp}$ and  $\mbox{Res}\,r_p = 2 k_{sp}/(\varepsilon_1+\varepsilon_2)$ (see \ref{GSP}). As a consequence,  Eq. (\ref{Gxxsp3}) takes the form,
\begin{eqnarray} \label{Gxxsp4}
 \hat{G}_{s,xx}(\mathbf{r}_B,\mathbf{r}_A)=i 
  \frac{ k_{sp}^3}{2 k_0^2 (\varepsilon_1+\varepsilon_2)}  H'_1(k_{sp}x_0) \,e^{- 2  k_{sp} z_0}.
\end{eqnarray}  
Taking the limit case for $\mathbf{r}_A$ tending to  $\mathbf{r}_B$, we obtain
\begin{eqnarray} \label{Gxxsp5}
 \hat{G}_{s,xx}(\mathbf{r}_B,\mathbf{r}_B)=i 
  \frac{ k_{sp}^3}{4 k_0^2 (\varepsilon_1+\varepsilon_2)}   \,e^{- 2  k_{sp} z_0}.
\end{eqnarray}  
Now, we can use Eq. (\ref{eq1B}) to calculate the force in the $x$ direction. 

Firstly, we begin by studying the case of normal incidence.  In the limit case in which  the diagonal elements of the effective  polarizability  tensor (\ref{alphaA}) are dominant \cite{NV2},  from Eq. (\ref{eq1B}), the $x$ component of the optical force on particle B can be approximated by,
\begin{eqnarray}\label{fplasmonx}
F^{(GSP,x)}_x(\mathbf{r}_B) = 
\frac{1}{2} \frac{k_0^2}{\varepsilon_0}\, \Bigg\{ \mbox{Re}\,
 p_{B x}^* \frac{\partial}{\partial x} [\hat{G}_{s,xx}(\mathbf{r},\mathbf{r}_A)] |_{\mathbf{r}_B} \, p_{A x}+\nonumber\\
  \mbox{Re}\,
 p_{B x}^* \frac{\partial}{\partial x} [\hat{G}_{s,xx}(\mathbf{r},\mathbf{r}_B)] |_{\mathbf{r}_B} \, p_{B x} \Bigg\}, 
\end{eqnarray}
for the case in which the incident electric field is along the $x$ axis, and 
\begin{eqnarray}\label{fplasmony}
F^{(GSP,y)}_x(\mathbf{r}_B) = 
\frac{1}{2} \frac{k_0^2}{\varepsilon_0}\,\Bigg\{ \mbox{Re}\,
 p_{B y}^* \frac{\partial}{\partial x} [\hat{G}_{s,y y}(\mathbf{r},\mathbf{r}_A)] |_{\mathbf{r}_B} \, p_{A y}+\nonumber\\
%\frac{1}{2} \frac{k_0^2}{\varepsilon_0}\, 
\mbox{Re}\,
 p_{B y}^* \frac{\partial}{\partial x} [\hat{G}_{s,yy}(\mathbf{r},\mathbf{r}_B)] |_{\mathbf{r}_B} \, p_{B y} \Bigg\} , 
\end{eqnarray}
for the case in which the incident electric field is along the  $y$ axis. Note that we have included the superscript $x$ and $y$ in Eqs. (\ref{fplasmonx}) and (\ref{fplasmony}) to indicate the direction of the incident electric field. Since, $\hat{G}_{s,xx}(\mathbf{r}',\mathbf{r}')$ given by Eq. (\ref{Gxxsp5}) is imaginary, it is follows that $\frac{\partial}{\partial x}\hat{G}_{s,xx}(\mathbf{r}',\mathbf{r}')$ is imaginary,  consequently the last term in Eqs. (\ref{fplasmonx}) and (\ref{fplasmony}) does not contribute to the force.  Taking  derivative respect to $x$ variable, and considering $p_{A x}=p_{B x}=p_x$ (the field is normally incident on the surface), Eq. (\ref{fplasmonx}) is written as
\begin{eqnarray}\label{fplasmonx2}
F^{(GSP,x)}_x(\mathbf{r}_B) = \mbox{Re} \Bigg\{ i
\frac{1}{2} \frac{k_0^2}{\varepsilon_0}\, 
 |p_{x}|^2  
  \frac{ k_{sp}^4}{2 k_0^2 (\varepsilon_1+\varepsilon_2)}  H''_1(k_{sp}x_0) \,e^{- 2  k_{sp} z_0}\Bigg\} \approx \nonumber\\
%  - \mbox{Re} 
%\frac{3i}{4 \varepsilon_0}\, 
% |p_{x}|^2  
%  \frac{ k_{sp}^4}{4  (\varepsilon_1+\varepsilon_2)}  H_1(k_{sp} x_0) \,e^{- 2  k_{sp} z_0} = \nonumber\\
  \frac{1}{4 \varepsilon_0}\, 
 |p_{x}|^2  
  \frac{ k_{sp}^4}{  (\varepsilon_1+\varepsilon_2)}  Y_1(k_{sp} x_0) \,e^{- 2  k_{sp} z_0}.
\end{eqnarray}
%where $H_1(z)=J_1(z)+iY_1(z)$.
In the last equality we have given the plasmonic force for large range distances, $x_0>>1/k_{sp}$.  
Following the same steps for obtaining Eq. (\ref{fplasmonx2}), from Eqs. (\ref{fplasmony}) and (\ref{Gyy}) with $\theta=0$, we obtain
\begin{eqnarray}\label{fplasmony2}
F^{(GSP,y)}_x(\mathbf{r}_B) \approx %\mbox{Re} i \frac{1}{2} \frac{k_0^2}{\varepsilon_0}\,  |p_{x}|^2     \frac{ k_{sp}^4}{2 k_0^2 (\varepsilon_1+\varepsilon_2)}  H''_1(k_{sp}x) \,e^{- 2  k_{sp} z'} =\nonumber\\  - \mbox{Re} \frac{i}{2 \varepsilon_0}\,  |p_{x}|^2    \frac{ k_{sp}^4}{4  (\varepsilon_1+\varepsilon_2)}  H_1(k_{sp}x) \,e^{- 2  k_{sp} z'} =\nonumber\\
  \frac{1}{4 \varepsilon_0}\, 
 |p_{y}|^2  
  \frac{ k_{sp}^4}{  (\varepsilon_1+\varepsilon_2)}  \frac{Y_2(k_{sp}x_0)}{k_{sp}x_0} \,e^{- 2  k_{sp} z_0},
\end{eqnarray}
where we have used $\frac{d}{dx}(H_1(x)/x)=-H_2(x)/x$. Equations  (\ref{fplasmonx2}) and (\ref{fplasmony2}) give the GSP optical force in $x$ direction between two particles aligned along the $x$ axis, when the field is normally incident with its electric field parallel to the $x$ axis (Eq. (\ref{fplasmonx2})) and parallel to the $y$ axis (Eq. (\ref{fplasmony2})).

Next, we obtain  the plasmon contribution for oblique incidence in the case for which the plane of incidence coincides with the $x-z$ plane.  
To do this, by following the same procedure as that used to obtain Eq. (\ref{Gxxsp4}), we extract the GSP contribution from all components of the Green tensor (\ref{G3p}) and replace these into Eq. (\ref{eq1B}), % For the case in which the incident  electric field is on the $x-z$ plane, the plasmon contribution to the optical force is written as,
\begin{eqnarray}\label{fplasmon_angulo}
F_x(\mathbf{r}_B) \approx \frac{1}{2}\mbox{Re}  \Bigg\{i k_0   \sin \theta_i 
 [p^*_{Bx} E_{0x}(\mathbf{r}_B)+p^*_{Bz} E_{0z}(\mathbf{r}_B)] \nonumber\\
 + \frac{k_{sp}^4}{4 \varepsilon_0 (\varepsilon_1+\varepsilon_2)}\, \Bigg(  -i H_1(k_{sp}x_0)[p_{Bx}^*p_{Ax} 
+ %p_{By}^* H_2(k_{sp}x_0)p_{Ay}+\nonumber\\
  p^*_{Bz}p_{Az}]\nonumber\\
  -i H_1'(k_{sp}x_0)[p_{Bz}^*p_{Ax}-p_{Bx}^*p_{Az}] 
  +\frac{i}{2}[p_{Bz}^*P_{Bx}-p_{Bx}^*P_{Bz}]\Bigg)\Bigg\}\nonumber\\
  \times e^{-2k_{sp}z_0},
  %\Bigg\{E_i i k_0   \sin(\theta_i) [p^*_{Bx} \cos(\theta_i)+p^*_{Bz} \sin(\theta_i)]\Bigg\}
\end{eqnarray}
where we have used that  $\hat{G}_{s,xy}(\mathbf{r}_B,\mathbf{r}_A)=\hat{G}_{s,yz}(\mathbf{r}_B,\mathbf{r}_A)=0$ because $\theta =0$ or $\pi$ in Eqs. (\ref{Gxy}) and (\ref{Gyz}) (see \ref{Green}). %This is consistent with the field distribution generated by a dipole $\mathbf{p}_A$ at the position of $\mathbf{p}_B$ \cite{OZC}. 
The first term in Eq. (\ref{fplasmon_angulo}) corresponds to the component force along the $x$ axis of the incident and reflected by the graphene plane surface  fields, $E_{0x}(\mathbf{r}_B)=E_i \cos \theta_i [e^{-i k_0 \cos \theta_i z_0}-r_p e^{i k_0 \cos \theta_i z_0}] e^{i k_0  \sin \theta_i x_B}$ and $E_{0z}(\mathbf{r}_B)=E_i \sin \theta_i [e^{-i k_0 \cos \theta_i z_0}+r_p e^{i k_0 \cos \theta_i z_0}] e^{i k_0  \sin \theta_i x_B}$. The second  term represents the interaction between two dipole moments, both oriented in the $x$ and $z$ axis, respectively, via surface plasmon  excitations. The third term represents the interaction between dipole $A$ and $B$, one of them is along the $x$ axis and the other is along  the $z$ axis, and the last term represents the interaction between both components, $x$ and $z$, of the same dipole moment. The force on particle A can be obtained from Eq. (\ref{fplasmon_angulo}) by permuting  the indices A and B and changing the sign of function $H_1(k_{sp}x_0)$ in the second term.

%From Eq. (\ref{fplasmonx2}) we see that the maximum contribution of GSPs to the force is given when 

%We closely follow a Green method developed in Ref. [3] and the Green integrals are rigorously solved modifying the original integration path and applying a gaussian quadrature as in Ref. [8].

\section{Results.}

In this section, we apply the sketch developed in  above sections to calculate the optical force between particles under plane wave illumination.  Firstly, our calculations are made for spherical particles and then we take care of others shapes different to spherical ones, like cylinders and cubes.  %for normal incidence ($\theta_i=0$) and then, we take care of the most general case of  oblique incidence. 
The relative permittivity of the particles are $\varepsilon_p=3.9$.   
The radius of the spheres is  $R=90$nm and the size of cubes an cylinders are taken in such a way the volumes of these three objects are equal. 
We use the large doping condition for which  the graphene conductivity takes a Drude behaviour (see  \ref{grafeno}). %In all the calculations we taken  $\mathbf{r}_A=z'\hat{z}$ and $\mathbf{r}_B=x'\hat{x}+z'\hat{z}$  the dimer axis is along the $x$ axis. 
\begin{figure}[htbp!]
    \centering
%    \subfloat[]
    {\includegraphics[width=0.45\textwidth]{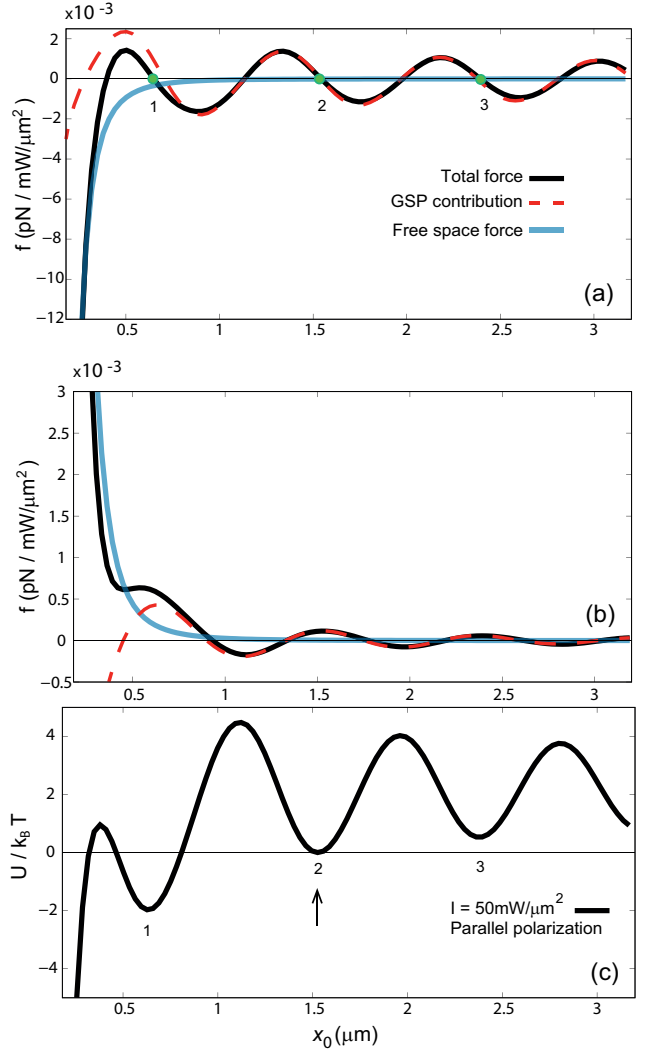}}
%    \hspace{1mm}
%    \subfloat[]{\includegraphics[width=0.45\textwidth]{OSA2.png}}
    \caption{Optical force as a  function of the interparticle distance $x_0$ for (a) parallel and (b) perpendicular  polarizations, and (c) potential energy binding normalized to $k_B$T for  parallel polarization. %The permittivity $\varepsilon_1=1$ and  $\varepsilon_2=\varepsilon_p=3.9$. 
    The potential energy values have been calculated for an intensity $I=50mW/\mu$m$^2$. The arrow in (c) indicates the position of zero potential, which coincides with the equilibrium position 2.  The frequency  $\omega/c=0.325\mu$m$^{-1}$,  the particle radius $R=90$nm and the distance $z_0=100$nm. The graphene parameters are $\mu_g=0.3$eV and $\gamma_g=0.1$meV.  %(a) Parallel polarization and (b) perpendicular polarization. Points $1,\,2,\,3$ and $1,\,2,\,3,\,4$ in (a) and (b) indicate the first stable points. 
    }
    \label{ob1}
\end{figure}
We have normalized the optical force $F$  with respect to the incident density of power $I=|E_0|^2/(2 Z_0)$, $Z_0=\sqrt{\mu_0/\varepsilon_0}$ is the vacuum impedance, \textit{i.e.},  we calculated  $f=F/I$, where $F$ is given in pico--Newton ($p$N) and $I$ is given in $mW/\mu$m$^2$. 

In all the cases, the optical force has been rigorously  calculated. To do this, the Green tensor components are calculated  following the same procedure as one applied in Ref. \cite{OZC} %. We surround the pole singularities by deforming the integration path into  two paths on the complex $k_{||}$ plane. The first of them is an elliptical path into the fourth quadrant and the second is a vertical path (parallel to imaginary $k_{||}$ axis). Then we used a 32 point Gauss Legendre quadrature to calculate the Green integrals (\ref{G3p}) on two paths 
(see \ref{Green_camino_de_integracion}). 

Figures \ref{ob1}a and \ref{ob1}b  show  the optical force for normal incidence as a function of the interparticle distance  $x_0$, rigorously calculated using Eq. (\ref{eq1B}), for   illumination parallel (electric field parallel to the incidence plane, $E_i=E_{||}$) and perpendicular (electric field perpendicular to the incidence plane,  $E_i=E_{\perp}$), respectively,  rigorously calculated using Eq. (\ref{eq1B}). 
The chemical potential of graphene and frequency are  $\mu_g=0.3$meV and  $\omega/c=0.325\mu$m$^{-1}$ (wavelength $\lambda=19.3\mu$m), respectively. %Since, for values of $x_0$ large enough, the features of the curve  for perpendicular polarization are qualitatively similar to those for parallel polarization, for clarity we have plotted Figure \ref{ob1}b in the range $x_0<1\mu$m.  In both cases, w
In both cases, we observe  that the force periodically varies with the $x_0$ coordinate with a period   $\approx 0.8\mu$m. At the incident field frequency, the real part of the calculated GSP propagation
constant is $\mbox{Re} \,k_{sp}=7.55\mu$m$^{-1}$,  corresponding   to a plasmon wavelength  $\lambda_{sp}=0.83\mu$m which agrees well with
the numerically calculated period of the force.

%SPPs are sustained and dominate the field in an
To gain insight about the role of GSPs in the interparticle force, in Figure \ref{ob1} we plotted the plasmon contributions given by Eqs. (\ref{fplasmonx2}) and (\ref{fplasmony2}). We observe  that for $x_0>1\mu$m these curves match with those rigorously calculated with Eq. (\ref{eq1B}),  indicating that  for interparticle distance  values large enough, surface plasmon excitations 
dominate the optical binding calculated in Figure \ref{ob1}.  
On the contrary, for small values of $x_0$ ($x_0<1\mu$m),  the surface plasmon contribution departs from the total optical force which is very approximately
equal to the one in free space, \textit{i.e.}, to the optical force in absence of graphene. In fact, in this  range, the force is attractive when the particles are illuminated with parallel polarization, whereas it is repulsive when particles are illuminated with perpendicular polarization.  
Moreover, in this range the optical force  changes rapidly with $x_0$. 

All these characteristics can be understood by comparing the contributions of free space and GSP modes in the short distance range between particles. %Since, in this range, the interaction can be described by 

\begin{equation}
F_x^{(FS)} \approx \frac{1}{2} \frac{k_0^2}{\varepsilon_0} \mbox{Re} \Bigg\{|p_{j}|^2 \frac{\partial}{\partial x_j'} G_{0,x_jx_j}(\mathbf{r}',\mathbf{r})\Bigg\},
\end{equation}
where $j=x$ for parallel polarization and $j=y$ for perpendicular polarization, $G_{0,xx}$ and $G_{0,yy}$ are the components of free Green tensor (Eq. (\ref{Glibre})),  %$\hat{\mathbf{G}}_0(\mathbf{r},\mathbf{r}')$, with    
$\mathbf{r}=x_0\hat{x}+z_0\hat{z}$ and   $\mathbf{r}'=z_0\hat{z}$. 
Taking the derivative of $G_{0,xx}$ and $G_{0,yy}$ components of free Green tensor (Eq. (\ref{Glibre})), we obtain,  $F_x^{(FS)} \approx - %3/(4 \pi  \varepsilon_0) 
|p_x|^2/x_0^4$ for parallel polarization and $F_x^{(FS)} \approx %3/(8 \pi \varepsilon_0) 
|p_y|^2 / x_0^4$ for perpendicular polarization. The negative sign indicates that the force is attractive for  parallel polarization, whereas the positive sign corresponds to a repulsive force for perpendicular polarization.  
This is consistent with the intuitive idea that  
%This is true because 
in this range, where the plasmonic contribution can be neglected,   the polarization charges on the surface of particles are asymmetric for parallel polarization, resulting in an attractive force. On the contrary, for perpendicular polarization the surface charge distribution on particles is symmetric, resulting in a repulsive force \cite{Salary}. 
%This is true because the particles are illuminated with an incident  electric field parallel to the axis joining them,  thus the polarization charges on the surface of particles are asymmetric, resulting in an attractive force \cite{Salary}. %On the contrary, for perpendicular polarization the surface charge distribution on particles is symmetric, resulting in a repulsive force \cite{Salary}. 
%are the source of 
%This fact comprises the first demonstration that, for interparticle distance  values large enough,  surface plasmon excitations %are the source of dominates the optical force observed in Figure \ref{figure1}. 
%the optical binding is mostly driven by the GSP contributions.
%
\begin{figure}[htbp!]
    \centering
%    \subfloat[]
    {\includegraphics[width=0.5\textwidth]{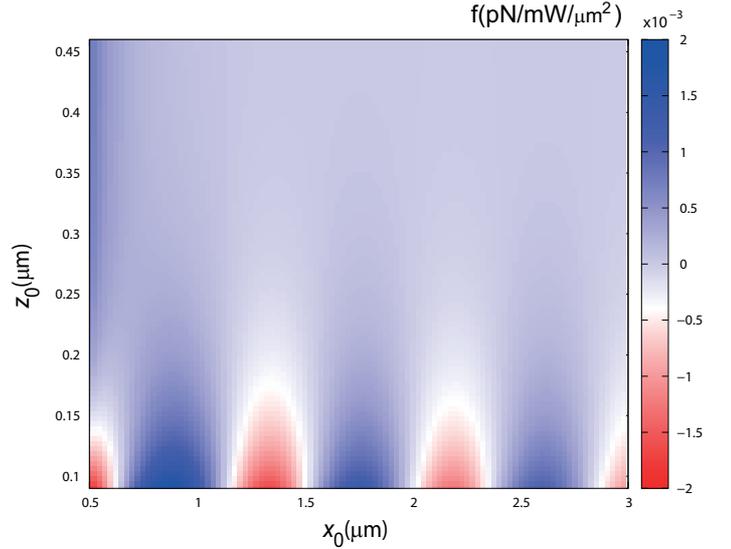}}
%    \hspace{1mm}
%    \subfloat[]{\includegraphics[width=0.45\textwidth]{OSA2.png}}
    \caption{Map of the optical force as a function of the interparticle distance $x_0$ and the height $z_0$ for  $\omega/c=0.325\mu$m$^{-1}$ and $\mu_g=0.3$eV. All other parameters are the same as in Figure \ref{ob1}.}
    \label{ob2}
\end{figure}
%
%On the other hand, 
While the spatial dependence of the free space contribution is $\approx \pm 1/x_0^4$, the GSP contribution given by Eqs. (\ref{fplasmonx2}) and  (\ref{fplasmony2}) is $F_x^{(GSP)}\approx 1/x_0^3$. From the above analysis, we conclude that  free space contributions dominate the optical force interaction at short distances. %  $x_0$ values small enough. 

%In addition, in this range the optical force  changes so rapidly with $x_0$. This behaviour can be understood as follows. Since, in this range, the interaction can be described by the free space Green function  $\hat{\mathbf{G}}_0(\mathbf{r},\mathbf{r}')$, with    $\mathbf{r}=x_0\hat{x}+z_0\hat{z}$ and   $\mathbf{r}'=z_0\hat{z}$, the optical force (\ref{eq1B}) is written as, 
%
%
%where $j=x$ for parallel polarization and $j=y$ for perpendicular polarization. Taking the derivative of $G_{0,xx}$ and $G_{0,yy}$ components of free Green tensor (Eq. (\ref{G3})), we obtain,  $F_x \approx - %3/(4 \pi  \varepsilon_0) |p_x|^2/x_0^4$ for parallel polarization and $F_x \approx %3/(8 \pi \varepsilon_0) |p_y|^2 / x_0^4$ for perpendicular polarization. 

%In the inset 
In Figure \ref{ob1}a  we have plotted points at stable positions, these are, points on the $x_0$ axis  where the optical force pass from positive to  negative values \cite{NV2}. In this range, the normalized total force reaches a value $\approx 1.4\mbox{x}10^{-3}p$N, \textit{i.e.}, a value of $0.07p$N for intensities of 50$mW/\mu$m$^2$. Taking into account that the period of the force is $\lambda_p=0.83\mu$m, the trapping potential of the optical binding approaches $4k_B$T at room temperature   and moderate intensities involved in optical trapping   \cite{NV2,kostina}.    This fact can be seen in Figure \ref{ob1}c, where  we have calculated the potential energy of the optical binding for parallel polarization  by integration of the force for an intensity of $50mW/\mu$m$^2$ (we take the potential energy as null for the equilibrium position 2). To do this, we used the fact that  the particles are  smaller than the wavelength,  which results
in low %and nonresonant polarizabilities so that where Im(αeff )  Re(αeff ), as Im(αeff )  (R6/λ3 ), and R  λ (see Fig. 2).
values of the imaginary part of the polarizability $\hat{\alpha}_j$ ($j=A,\,B$) and, consequently, the  non conservative part of the force (radiation force) can be neglected \cite{NV2,Novotny}.     
On the other hand, note that  stable first positions are at $x_0 \approx 1.5\mu$m, a value near 15 times lower  than that corresponding to the free space, which falls at $x_0\approx 22\mu$m [not shown in Figure \ref{ob1}], %the photon wavelength, %
which demonstrates the high subwavelength binding that is  provided by GSPs.  

%In the inset in Figure \ref{ob1}a and in Figure \ref{ob1}b  we have plotted points at stable positions, these are, points on the $x_0$ axis  where the optical force pass from positive to  negative values \cite{NV2}. It is worth noting that, for parallel polarization, stable first positions (see inset in Figure \ref{ob1}a) are at $x_0 \approx 0.4\mu$m, a value near 30 times lower  than that corresponding to the free space, which falls at $x_0\approx 13\mu$m (see Figure \ref{ob1}a). Similarly,   for perpendicular polarization, the first stable position falls near $x_0 \approx 0.25\mu$m which is even lower than that for parallel polarization. These features demonstrate the high subwavelength binding that is  provided by GSPs.

%Since the optical binding characteristics for  perpendicular polarization are qualitatively similar to those for parallel polarization, in what follows, we only present calculations for parallel polarization.  

By comparing Figures \ref{ob1}a and \ref{ob1}b, we observe that for $x>1\mu$m, the force values for perpendicular polarization are one order of magnitude less than those for parallel polarization. This is because, in this range, where the force is dominated by the plasmon contribution, the optical force % given by  Eq. (\ref{fplasmony2}) 
for perpendicular polarization decays faster than that corresponding to parallel polarization ($F_x^{(y)} \approx Y_2(k_{sp}x_0)/(k_{sp} x_0)$ and $F_x^{(x)} \approx Y_1(k_{sp}x_0)$). % as  $F_x^{(y)} \approx Y_2(k_{sp}x_0)/(k_{sp} x_0)$ with the interparticle distance and Eq. (\ref{fplasmony2}) for perpendicular polarization. While Eq. (\ref{fplasmonx2}) provides an interparticle  dependence of the force as $F_x^{(x)} \approx Y_1(k_{sp}x_0)$, Eq. (\ref{fplasmony2}) provides a interparticle dependence as $F_x^{(y)} \approx Y_2(k_{sp}x_0)/(k_{sp} x_0)$, which explains the greater decay of the force for perpendicular polarization compared with that for parallel polarization.  
As a consequence, the stable trapping for moderate intensity, with  the parameters considered here, is then impossible for perpendicular polarization.

In what follows, we only present calculations for parallel polarization.  

In   Figure \ref{ob2} we calculated the optical force for normal incidence as a function of $x_0$ and $z_0$ for $\mu_g=0.3$eV and $\omega/c=0.325\mu$m$^{-1}$. 
We see that the force presents the same periodic behaviour, with respect $x_0$ variable,  as in Figure \ref{ob1}, passing from positive to negative values with the period of the plasmon  wavelength $\lambda_{sp}=0.83\mu$m. 
In addition, the force values quickly decrease with the height $z_0$ and practically disappear for $z_0$ values larger than $0.35\mu$m. This can be understood from Eqs. (\ref{fplasmonx2})  and (\ref{fplasmony2}), where we observe   an exponential decay of the force with $z_0$ as $\approx \exp(-2 k_{sp} z_0 )$. 

\begin{figure}[htbp]
    \centering
%    \subfloat[]
%    {\includegraphics[width=0.5\textwidth]{rho_vs_omegac3c.eps}}
    {\includegraphics[width=0.45\textwidth]{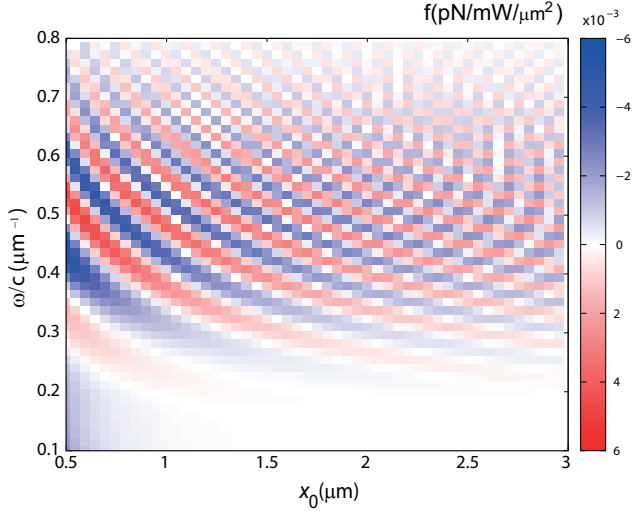}}
%    \hspace{1mm}
%    \subfloat[]{\includegraphics[width=0.45\textwidth]{OSA2.png}}
    \caption{Map of the optical force as a function of the interparticle distance and frequency. All other  parameters are the same as in Figure \ref{ob1}.}
    \label{ob3}
\end{figure}

Next, we study the frequency dependence of the optical binding. Figure \ref{ob3} shows a map of the optical force as a function of $\omega/c$ frequency and the interparticle distance $x_0$ for $\mu_g=0.3$eV. % and parallel polarization. 
For a fixed frequency value, we observe a periodic behaviour together with a  decay of the oscillation amplitude with the $x_0$ distance. This period decreases with frequency,  which is in agreement with the fact that the GSP  propagation constant, given by Eq. (\ref{sp_cuasi}), is an  increasing function of frequency. It is worth noting that, although  the force intensity decreases with the interparticle distance, the dependence with frequency is not like this.  For a fixed $x_0$ value,  the force intensity increases until it reaches the maximum value at $\approx 0.5\mu$m$^{-1}$ and then, the force intensity monotonously  decreases with frequency. This behaviour can be understood from Eq. (\ref{fplasmonx2}),  where we can see that the force depends on the GSP propagation constant as $F_x^{(x)}(x_0) \approx k_{sp}^4 Y_1(k_{sp}x_0) e^{-2 k_{sp}z_0}$. Then, for a fixed value of $x_0$, the function $F_x^{(x)}$ reaches its  maximum value for $k_{sp}=2/z_0=20\mu$m$^{-1}$. From the dispersion relation plotted in Figure \ref{alfasp} for $\mu_g=0.3$eV, we find that the corresponding value for $k_{sp}(\omega)=20\mu$m$^{-1}$ is $\omega/c=0.52\mu$m$^{-1}$, a value that agrees well with that numerically found.  %We have observed (not shown here)  that, for perpendicular polarization, the force present a qualitatively similar map to that for parallel polarization shown in Figure \ref{ob3}. 
Even though for perpendicular polarization, the force presents a similar behaviour to that for parallel polarization shown in Figure \ref{ob3},  it is worth noting that, from Eq. (\ref{fplasmony2}), the frequency value where the force for perpendicular polarization reaches the maximum value verify the condition $k_{sp}(\omega)=3/(2 z_0)=15\mu$m$^{-1}$, leading to $\omega/c=0.46\mu$m$^{-1}$.

%fades quickly
%
\begin{figure}[htbp]
    \centering
%    \subfloat[]
    {\includegraphics[width=0.5\textwidth]{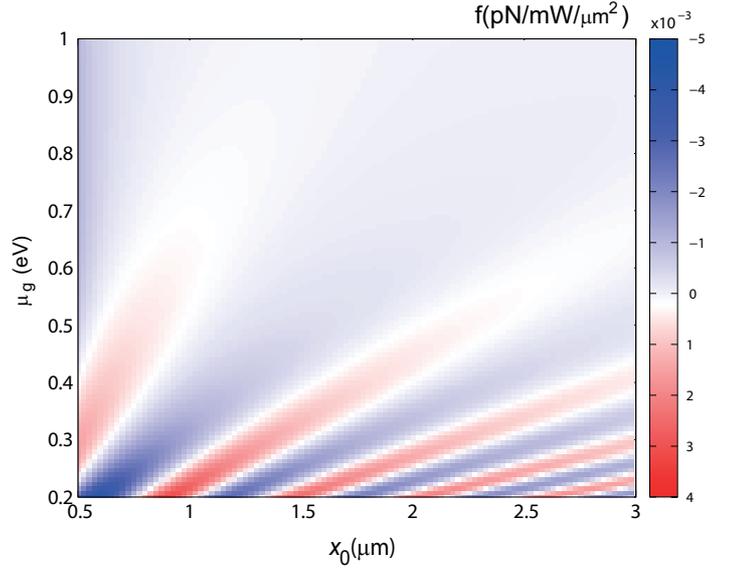}}
%    \hspace{1mm}
%    \subfloat[]{\includegraphics[width=0.45\textwidth]{OSA2.png}}
    \caption{Map of the optical force as a function of the interparticle distance and chemical potential. All parameters are the same as in Figure \ref{ob1}.}
    \label{ob4}
\end{figure}

In order to  study the optical binding 
dependence with graphene parameters,  we calculate the interparticle force dependence with the chemical potential for normal incidence. Figure \ref{ob4} shows  the optical force as a function of the $x_0$ distance and the chemical potential. % for parallel polarization. 
%Results for perpendicular polarization (not shown here) show the same behaviour.  %(a) and perpendicular (b) polarizations. 
We observe zones for which the force is positive (red zones) and others where the force is negative (blue zones). Note that for a fixed chemical potential value, the force is a periodic function on the interparticle distance $x_0$, and that this period depends on the chemical potential value, being  an increasing function of $\mu_g$. This can be understood from the quasistatic expression for the GSP propagation constant. From Eq. (\ref{sp_cuasi}), we see that $k_{sp}$ is a decreasing function of $\mu_g$, thus the period of the function $Y_1(k_{sp}x_0)$  in Eq. (\ref{fplasmonx2}) and $Y_2(k_{sp}x_0)$ in Eq. (\ref{fplasmony2}) increases with $\mu_g$. Unlike the case of surface plasmons on metallic materials, for which the plasmon kinematic quantities are fixed by an  invariable charge density, GSPs provide a dynamic control on the optical binding period, and consequently, on the stable interparticle distance by varying the chemical potential of graphene. 

\begin{figure}[htbp]
    \centering
%    \subfloat[]
    {\includegraphics[width=0.45\textwidth]{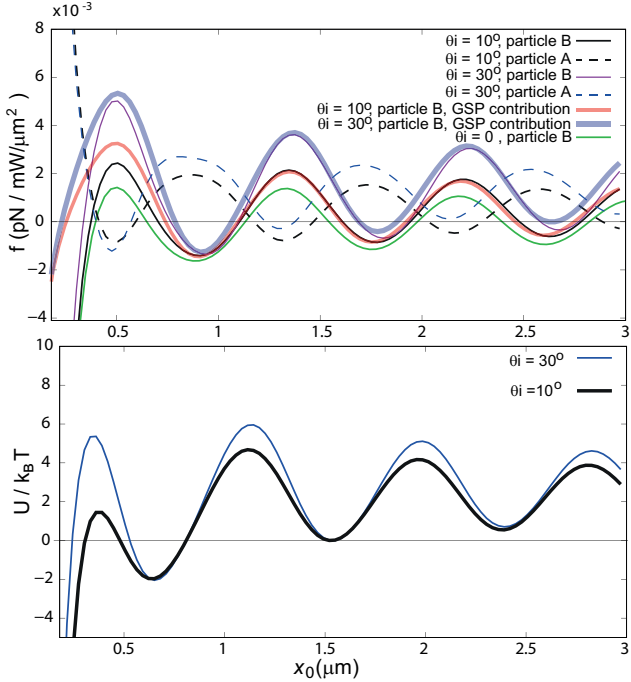}}
%    \hspace{1mm}
%    \subfloat[]{\includegraphics[width=0.45\textwidth]{OSA2.png}}
    \caption{(a) Optical force and (b) potential energy of interaction between the particles A and B normalized to $k_B$T,  as functions of the interparticle distance $x_0$ for $\omega/c=0.325\mu$m$^{-1}$ The incidence angles are $\theta_i=10^\circ,\,30^\circ$. The curve for normal incidence is given as a reference.  
    The potential energy values have been calculated for an intensity $I=50mW/\mu$m$^2$. All parameters are the same as in Figure \ref{ob1}.}
    \label{incidencia_oblicua}
\end{figure}

Now, we investigate the optical binding properties under oblique  incidence. We focus on the case in which the plane of incidence coincides with the $x-z$ plane. Figure \ref{incidencia_oblicua} shows the optical force, rigorously calculated, as a function of $x_0$ for %various values of  
 $\theta_i=10^\circ\,\mbox{and}\,30^\circ$. The optical force on particle B for normal incidence is plotted as a reference.    
 From  Figure \ref{incidencia_oblicua}a we observe that the force on particle A is not opposite to the force on particle B, as in the case of normal incidence. This effect is essentially caused by two mechanisms.  The first mechanism is associated with the incoming photon momentum parallel to the graphene surface: both particles are pushed in the $+x$ direction, \textit{i.e.}, in the direction of the incident photon  propagation constant parallel to the graphene surface. The second mechanism is associated with the break of the mirror symmetry imposed by the oblique incidence,  giving rise  to different dipole moments induced  on particles A and B and different forces on each of the particles take place. As a consequence, the equilibrium positions  obtained for particle B (positions where the particle B is fixed) do not coincide with those obtained for particle A. In addition, as $\theta_i$  increases, the equilibrium positions are lost, as we can see % particles  
 for $\theta_i=30^\circ$, where the particle A does not reaches any equilibrium position for  $x_0>1.5\mu$m. On the other hand, the potential energy of interaction between particles, \textit{i.e.}, the potential referred to the relative positions between particles,  reaches minimum values, as can be seen in Figure \ref{incidencia_oblicua}b where we have computed the potential for $\theta_i=10^\circ$, $30^\circ$ and  an intensity $I=50mW/\mu$m$^2$.  We recall that  these positions   do not correspond to  stable equilibrium in which both  particles are  fixed, but to positions  where the system attains internal equilibrium.  A similar behaviour has been found  for a dielectric interface in the  total reflection configuration \cite{NV1}. By using Eq. (\ref{fplasmon_angulo}), we  have  calculated (Figure \ref{incidencia_oblicua}a) the GSP contribution to the optical force on particle B for $\theta_i=10^\circ$ and $30^\circ$. As in case of normal incidence, for an interparticle distance large enough, the plasmonic contribution follows well the behaviour of the optical force rigorously calculated by using Eq. (\ref{eq1B}).

Finally, we explore the influence of the shape of particles on the optical binding. To do this, cubic (with edge length $a$ and cylindrical (with height $h$ equal to diameter $d$)  shapes  have been considered. The values of $a$ and $h$ have been chosen in such a way the volumes of the  cubic, cylinder and sphere coincide.  The quasistatic polarizabilities  have been calculated  using the formulas presented in   \cite{cubo_cilindro}. 

\begin{figure}[htbp]
    \centering
%    \subfloat[]
    {\includegraphics[width=0.45\textwidth]{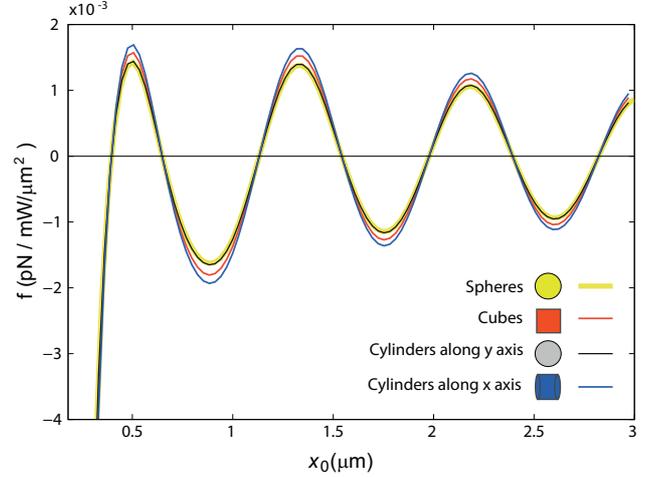}}
%    \hspace{1mm}
%    \subfloat[]{\includegraphics[width=0.45\textwidth]{OSA2.png}}
    \caption{Optical force as a function of the interparticle distance $x_0$ for normal incidence and various shapes. The radius of the sphere is  $R=90$nm and the volume of cubes and cylinders are equal to that of the sphere. All other parameters are the same as in Figure \ref{ob1}.}
    \label{figura7}
\end{figure}

From Figure \ref{figura7} we see that the  optical force curve   for the case of two cylinders placed with the cross section along $x-z$ plane (cylinders along y axis) matches with that corresponding to spherical particles. This is to be expected because the value of the polarizability  along an axis contained in the cylinder cross section almost coincides with that of the sphere of the same volume for $\varepsilon_{p}=3.9$ \cite{cubo_cilindro}. 
When the cylinders are placed with their symmetry axis along the incident electric field direction (cylinders along $x$ axis),  the force amplitude increases as a consequence of the polarizability increment along this direction. On the other hand, the force curve  for a cubic shape falls between the  curves corresponding to spheres and cylinders aligned along the $x$ axis.

%This is consistent with the fact that the polarizability along the cylinder  symmetry axis is larger than that  along the cross section. 

\section{Conclusions} 

In conclusion, we have analyzed the benefits of a  graphene sheet as a  nanoparticle binder. Due to super confinement of GSPs, we found equilibrium interparticle distances of few micrometers for low frequency plane wave  incidence, THz and IR.  Moreover,  these equilibrium distances can be dynamically controlled  by chemical potential variations, a fact that highlights the use of graphene for optical binding applications. 

In addition, using the dispersion properties of surface plasmons on a   perfectly flat and infinite graphene sheet,  we have obtained an analytical  expression for the binding force that quantitatively explains the main results obtained by applying the rigorous  Green's method.

%The objective of the present work is to   Even though by incrementing  frequency (or tuning the chemical potential) we can increses the optical force to achieve greater values of the potential well depth, it is not the objective of the present work to optimize the parameters of the system in order to maximize the optical binding effects. 

The possibility to incorporate another  graphene sheet forming a graphene parallel waveguide \cite{CuevasJOpt} or by creating a space layer forming an attenuated total reflection structure \cite{CuevasPRA}, allows  interesting degree of freedoms  to increase the GSP field and, as a consequence,  the depth of the potential well of the optical force between particles. Although we are planning to report the results of such studies in future papers, as a first step, here we have restricted ourselves to performing an analysis of the optical binding properties on a single graphene sheet. In this way, we believe that our results are valuable in the framework of optical binding  including graphene plasmonic  structures.

\section*{Acknowledgment}
The authors acknowledge the financial supports of Universidad Austral O04-INV00020 and Consejo Nacional de Investigaciones Científicas y Técnicas (CONICET).

\appendix

\section{Graphene conductivity} \label{grafeno}
\setcounter{equation}{0}
\renewcommand{\theequation}{A{\arabic{equation}}}
We consider the  graphene layer as an infinitesimally thin, local and isotropic two--sided layer with frequency--dependent surface conductivity $\sigma(\omega)$ given by the Kubo formula \cite{milkhailov} , which can be read as  $\sigma_{loc}= \sigma^{intra}+\sigma^{inter}$, with the intraband and interband contributions being
\begin{equation} \label{intra}
\sigma^{intra}(\omega)= \frac{2i e^2 k_B T}{\pi \hbar^2 (\omega+i\gamma_g)} \mbox{ln}\left[2 \mbox{cosh}(\mu_g/2 k_B T)\right],
\end{equation}  
\begin{eqnarray} \label{inter}
\sigma^{inter}(\omega)= \frac{e^2}{\hbar} \Bigg\{   \frac{1}{2}+\frac{1}{\pi}\mbox{arctan}\left[(\hbar \omega-2\mu_g)/2k_BT\right]-\nonumber\\
\frac{i}{2\pi}\mbox{ln}\left[\frac{(\hbar \omega+2\mu_g)^2}{(\hbar \omega-2\mu_g)^2+(2k_BT)^2}\right] \Bigg\},
\end{eqnarray}  
where $\mu_g$ is the chemical potential (controlled with the help of a gate voltage), $\gamma_g$ the carriers scattering rate, $e$ the electron charge, $k_B$ the Boltzmann constant and $\hbar$ the reduced Planck constant. For chemical potential values substantially larger than the thermal energy, \textit{i.e.},  $k_B T<<\mu_g$, the graphene conductivity can be well approached  by the intraband term which takes the Drude form,
\begin{equation} \label{intra2}
\sigma(\omega)= \frac{i e^2 \mu_g}{\pi \hbar^2 (\omega+i\gamma_g)}.
\end{equation}  

\section{Green tensor} \label{Green}
\setcounter{equation}{0}
\renewcommand{\theequation}{B{\arabic{equation}}}

The Green tensor $\mathbf{\hat{G}}(\mathbf{r},\mathbf{r}')$ is defined by the electric
field %$\mathbf{E(\mathbf{r})}$ 
at point $\mathbf{r}$ generated by an  electric dipole $\mathbf{p}$ located at the source
point $\mathbf{r}'$ and  %The Green tensor 
satisfies \cite{Novotny},
\begin{eqnarray} \label{G1}
\nabla \times \nabla \times \mathbf{\hat{G}}(\mathbf{r},\mathbf{r}')- k_0^2 \mathbf{\hat{G}}(\mathbf{r},\mathbf{r}')= \mathbf{I} \delta(\mathbf{r}-\mathbf{r}')),
\end{eqnarray}  
where $\mathbf{I}$ is the unit  tensor. By applying the superposition method, the solution of Eq. (\ref{G1}) can be expressed as a sum of two parts, one of them, $\mathbf{\hat{G}}_0(\mathbf{r},\mathbf{r}')$, is
associated to the primary dipole emission of the source   and the other, $\mathbf{\hat{G}}_s(\mathbf{r},\mathbf{r}')$,  takes into account the field scattered on $z=0$ surface. For $z,\,z'>0$ (upper half space), the expression for these functions have the form \cite{Novotny}: 
%For the vacuum half space, $\mathbf{z},\,\mathbf{z}'>0$,  $\mathbf{G}(\mathbf{r},\mathbf{r}')$
%
\begin{eqnarray} \label{G2}
 \mathbf{\hat{G}}_0(\mathbf{r},\mathbf{r}')=\frac{e^{i k_0 d}}{4 \pi d} \nonumber\\
\times \Bigg\{ [1 + \frac{i k_0 d-1}{k_0^2 d^2} ]  \mathbf{I}+ \frac{3-i3 k_0 d-k_0^2 d^2}{3 k_0^2 d^2} \hat{r}\hat{r} \Bigg\},
\end{eqnarray}  
where $d=|\mathbf{r}-\mathbf{r}'|$, 
 and  
\begin{eqnarray} \label{G3}
 \mathbf{\hat{G}}_s(\mathbf{r},\mathbf{r}')=\frac{i}{8 \pi^2} 
 \int_0^\infty dk_{||} \, \mathbf{f}(k_{||},\rho)\,e^{i  \gamma^{(1)}(z+z')},
\end{eqnarray}  
where $k_{||}$ is the wave vector parallel to the surface,  $\rho=\sqrt{d^2-(z-z')^2}$,  $\gamma^{(j)}=\sqrt{k_0^2\varepsilon_j-k_{||}^2}$ ($j=1,\,2$). The cartesian components of $\mathbf{f}(k_{||},\rho)$ are:  
\begin{eqnarray} \label{Gxx}
 \mathbf{f}_{xx}(k_{||},\rho)=2\pi \nonumber\\
 \times \Bigg\{ \frac{k_{||}}{\gamma^{(1)}} [-\cos(2\theta) \frac{J_1(k_{||}\rho)}{k_{||}\rho}+\cos^2(\theta) J_0(k_{||}\rho) ] r_s(k_{||})\nonumber\\
 + \frac{k_{||} \gamma^{(1)}}{k_0^2}  [\cos(2\theta) \frac{J_1(k_{||}\rho)}{k_{||}\rho}-\cos^2(\theta) J_0(k_{||}\rho) ] r_p(k_{||}) \Bigg\},
\end{eqnarray}  
\begin{eqnarray} \label{Gyy}
 \mathbf{f}_{yy}(k_{||},\rho)=2\pi \nonumber\\
\times \Bigg\{ \frac{k_{||}}{\gamma^{(1)}} [-\cos(2\theta) \frac{J_1(k_{||}\rho)}{k_{||}\rho}+\cos^2(\theta) J_0(k_{||}\rho) ] r_s(k_{||})] \nonumber\\
 - \frac{k_{||} \gamma^{(1)}}{k_0^2}  [\cos(2\theta) \frac{J_1(k_{||}\rho)}{k_{||}\rho}+\sin^2(\theta) J_0(k_{||}\rho) ] r_p(k_{||}) \Bigg\},
\end{eqnarray}  
\begin{eqnarray} \label{Gzz}
 \mathbf{f}_{zz}(k_{||},\rho)=2\pi \frac{k_{||}^3}{k_0^2 \gamma^{(1)}}  J_0(k_{||}\rho), 
\end{eqnarray}  
\begin{eqnarray} \label{Gxy}
 \mathbf{f}_{xy}(k_{||},\rho)=\mathbf{f}_{yx}(k_{||},\rho) = \nonumber\\
 \pi \Bigg\{  \frac{k_{||}}{\gamma^{(1)}}  r_s(k_{||})+  \frac{k_{||} \gamma^{(1)}}{k_0^2}   r_p(k_{||}) \Bigg\} J_2(k_{||}\rho) \sin(2\theta),
\end{eqnarray}  
\begin{eqnarray} \label{Gxz}
 \mathbf{f}_{xz}(k_{||},\rho)=-\mathbf{f}_{zx}(k_{||},\rho) = %\nonumber\\
 -2\pi i \frac{k_{||}^2}{k_0^2} J_1(k_{||}\rho) r_p(k_{||})  \cos(\theta),
\end{eqnarray}  
\begin{eqnarray} \label{Gyz}
 \mathbf{f}_{yz}(k_{||},\rho)=-\mathbf{f}_{zy}(k_{||},\rho) = %\nonumber\\
 -2\pi i \frac{k_{||}^2}{k_0^2} J_1(k_{||}\rho) r_p(k_{||})  \sin(\theta),
\end{eqnarray}  
where $\theta$ is the angle between the axis joining the particles and the $x$ axis, and $J_n(x)$ is the Bessel function of $n$th order. 
The complex amplitude
\begin{eqnarray}\label{fresnelp}
r_p=\frac{\frac{\gamma^{(1)}}{\varepsilon_1}-\frac{\gamma^{(2)}}{\varepsilon_2}+\frac{Z_0 \sigma}{k_0} \frac{\gamma^{(1)}}{\varepsilon_i} \frac{\gamma^{(2)}}{\varepsilon_2} }{\frac{\gamma^{(1)}}{\varepsilon_1}+\frac{\gamma^{(2)}}{\varepsilon_2}+\frac{Z_0 \sigma}{k_0} \frac{\gamma^{(1)}}{\varepsilon_1} \frac{\gamma^{(2)}}{\varepsilon_2}},
\end{eqnarray}
is the Fresnel reflection  coefficient for $p$ polarization (magnetic field parallel to the $z=0$ surface), $Z_0=\sqrt{\mu_0/\varepsilon_0}$ is the vacuum impedance, whereas 
\begin{eqnarray}\label{fresnels}
r_s=\frac{\gamma^{(1)}-\gamma^{(2)}- Z_0 k_0  \sigma  }{\gamma^{(1)}+\gamma^{(2)}+  Z_0  k_0 \sigma  },
\end{eqnarray}
is the Fresnel reflection for $s$ polarization (electric field parallel to the  $z=0$ surface). 

In the case in which the particles are along the $x$ axis,  at  heights  $z=z'=z_0$ and separated one from the other a distance $x_0$, \textit{i.e.},   $\mathbf{r}=x_0\hat{x}+z_0\hat{z}$ and $\mathbf{r}'=z_0\hat{z}$, we must take $\theta=0$ in expressions  (\ref{Gxx}) to (\ref{Gyz}). Moreover, Eq. (\ref{G2}) is written as,
\begin{eqnarray} \label{Glibre}
 \mathbf{\hat{G}}_0(\mathbf{r},\mathbf{r}')=\frac{e^{i k_0 x_0}}{4 \pi x_0} \nonumber\\
\times \Bigg\{ [1 + \frac{i k_0 x_0-1}{k_0^2 x_0^2} ]  \mathbf{I}+ \frac{3-i3 k_0 x_0-k_0^2 x_0^2}{3 k_0^2 x_0^2} \hat{x}\hat{x} \Bigg\}.
\end{eqnarray}  
% 

%The complex coefficients At(m) and Bt(m) in equations (8)–(10) correspond to the amplitude of upgoing (+z propagation direction) and downgoing (-z propagation direction) plane waves, respectively, and they are solutions of Helmholtz equation, whereas the former term in equation (8) is associated to the primary dipole emission of the source. There are two types of boundary conditions which must fulfill the solutions given by equations (7)–(10), boundary conditions at z = ¥ and boundary conditions at interfaces z=0 and z=d. The former requires either outgoing waves at infinity or exponentially decaying waves at infinity, depending on the values of α, β and ω.

%We consider that particles are aligned on the $x$ axis, \textit{i.e.}, $\theta=0$ in Eqs. (\ref{Gxx}) to (\ref{Gyz}).  

\section{Incident field and induced dipole moments} \label{Induced polarizability}
\setcounter{equation}{0}
\renewcommand{\theequation}{D{\arabic{equation}}}

To obtain the induced %polarizability  
dipole moments due to the incoming and scattered electric fields, we follows the same steps as in Ref. \cite{NV2}. In this manner, the dipole moments are written as
\begin{equation}\label{pA}
\mathbf{p}_A = \hat{\alpha}_s [ \mathbf{E}_0(\mathbf{x}_A) + \frac{k_0^2}{\varepsilon_0} \hat{\mathbf{\hat{G}}}(\mathbf{x}_A,\mathbf{x}_B) \mathbf{p}_B ], 
\end{equation}
%
%An expression for the dipole moment on particle B is obtained from Eq. (\ref{pA}) by changing A by B,
and
\begin{equation}\label{pB}
\mathbf{p}_B = \hat{\alpha}_s [ \mathbf{E}_0(\mathbf{x}_B) + \frac{k_0^2}{\varepsilon_0} \hat{\mathbf{G}}(\mathbf{x}_B,\mathbf{x}_A) \mathbf{p}_A ], 
\end{equation}
for particles A and B respectively. 
The first term in Eq. (\ref{pA}) (Eq. (\ref{pB})) represents the incident and reflected electric field, $\mathbf{E}_0(\mathbf{r})=\mathbf{E}_i(\mathbf{r})+\mathbf{E}_r(\mathbf{r})$, whereas the second term represents the field in particle A (B)  which is scattered by particle B (A). Here, $\hat{\alpha}_{s}$ is the particle polarizability that account the corrections due to the  scattering with the surface,
\begin{equation}\label{alpha_s}
\hat{\alpha}_s=\frac{\alpha_0} {\hat{\mathbf{I}}-\frac{k_0^2}{\varepsilon_0} \alpha_0 \hat{\mathbf{G}}_s(\mathbf{r}_j,\mathbf{r}_j ) },
\end{equation}
where $j=A,\,B$. In Eqs. (\ref{pA}) and (\ref{pB}) we have taking into account that particles A and B are identical. By solving Eqs. (\ref{pA}) and (\ref{pB}), we obtain
\begin{equation}\label{pA2}
\begin{array}{ll}
\mathbf{p}_A=\hat{\alpha}_s \frac{\mathbf{E}_0(\mathbf{x}_A)+\frac{k_0^2}{\varepsilon_0} \hat{\mathbf{G}}(\mathbf{x}_A,\mathbf{x}_B) \hat{\alpha}_s \mathbf{E}_0(\mathbf{x}_B)} {\hat{\mathbf{I}}-\frac{k_0^4}{\varepsilon_0^2} \hat{\alpha}_s\hat{\mathbf{G}}(\mathbf{x}_A,\mathbf{x}_B)\hat{\alpha}_s\hat{\mathbf{G}}(\mathbf{x}_B,\mathbf{x}_A)}.% =\\
%\hat{\alpha}_s \frac{\hat{\mathbf{I}}+\frac{k_0^2}{\varepsilon_0} \hat{\mathbf{G}}(\mathbf{x}_A,\mathbf{x}_B) \hat{\alpha}_s } {\hat{\mathbf{I}}-\frac{k_0^4}{\varepsilon_0^2} \hat{\alpha}_s\hat{\mathbf{G}}(\mathbf{x}_A,\mathbf{x}_B)\hat{\alpha}_s\hat{\mathbf{G}}(\mathbf{x}_B,\mathbf{x}_A)} \mathbf{E}_0,
\end{array}
\end{equation}
%
%where, in the last equality, we have taking into account that $\mathbf{E}_0(\mathbf{x}_A)=\mathbf{E}_0(\mathbf{x}_B)=\mathbf{E}_0$ for normal incidence. 
In case where the incidence is on the $x-z$ plane and for $p$ polarization (electric field on the incidence plane),  
the explicit form of the incident electric field $\mathbf{E}_i(\mathbf{r}_j)$ ($j=A,\,B$) is given by, 
\begin{equation}\label{Ei}
\begin{array}{ll}
\mathbf{E}_i(\mathbf{r}_A)=E_i [\,\cos\theta_i\,\cos\phi_i \hat{x}+\\ 
\,\cos\theta_i\,\sin\phi_i \hat{y}+  \sin\theta_i\hat{z}]\,e^{-ik_0\cos\theta_i\,z_0},\\
\mathbf{E}_i(\mathbf{r}_B)=E_i [\,\cos\theta_i\,\cos\phi_i\hat{x}+\\ 
\,\cos\theta_i\,\sin\phi_i\hat{y}+  \sin\theta_i\hat{z}]\,e^{ik_0(\sin\theta_i\,x_0-\cos\theta_i\,z_0)},
\end{array}
\end{equation}
where  $E_i$ is the amplitude of the incident plane     electric field. The explicit form of the reflected field $\mathbf{E}_r(\mathbf{r}_j)$ ($j=A,\,B$) can be obtained from Eq. (\ref{Ei}) using the reflection coefficients $r_p(k_{||})$ defined in Eqs. (\ref{fresnelp}). 

In many cases, for normal incidence for example, the electric fields at both particle positions are equals, \textit{i.e.}, $\mathbf{E}_0(\mathbf{r}_A)=\mathbf{E}_0(\mathbf{r}_B)$. In such cases, Eq. (\ref{pA2}) can be written as 
\begin{equation}\label{pA3}
\begin{array}{ll}
\mathbf{p}_A=%\hat{\alpha}_s \frac{\mathbf{E}_0(\mathbf{x}_A)+\frac{k_0^2}{\varepsilon_0} \hat{\mathbf{G}}(\mathbf{x}_A,\mathbf{x}_B) \hat{\alpha}_s \mathbf{E}_0(\mathbf{x}_B)} {\hat{\mathbf{I}}-\frac{k_0^4}{\varepsilon_0^2} \hat{\alpha}_s\hat{\mathbf{G}}(\mathbf{x}_A,\mathbf{x}_B)\hat{\alpha}_s\hat{\mathbf{G}}(\mathbf{x}_B,\mathbf{x}_A)}.% =\\
\hat{\alpha}_A %\frac{\hat{\mathbf{I}}+\frac{k_0^2}{\varepsilon_0} \hat{\mathbf{G}}(\mathbf{x}_A,\mathbf{x}_B) \hat{\alpha}_s } {\hat{\mathbf{I}}-\frac{k_0^4}{\varepsilon_0^2} \hat{\alpha}_s\hat{\mathbf{G}}(\mathbf{x}_A,\mathbf{x}_B)\hat{\alpha}_s\hat{\mathbf{G}}(\mathbf{x}_B,\mathbf{x}_A)} 
\mathbf{E}_0,
\end{array}
\end{equation}
where we have defined the effective polarizability for particle A as
%
%From Eq. (\ref{pA2}), the effective polarizability for particle A is defined as,
%
\begin{equation}\label{alphaA}
\begin{array}{ll}
\hat{\alpha}_A=\hat{\alpha}_s \frac{\hat{\mathbf{I}}+\frac{k_0^2}{\varepsilon_0} \hat{\mathbf{G}}(\mathbf{x}_A,\mathbf{x}_B) \hat{\alpha}_s } {\hat{\mathbf{I}}-\frac{k_0^4}{\varepsilon_0^2} \hat{\alpha}_s\hat{\mathbf{G}}(\mathbf{x}_A,\mathbf{x}_B)\hat{\alpha}_s\hat{\mathbf{G}}(\mathbf{x}_B,\mathbf{x}_A)}.
\end{array}
\end{equation}
Expressions for the dipole moment and polatizability on particle B are obtained from Eq. (\ref{pA2}) and (\ref{alphaA}), respectively, by interchanging A by B. 

\section{Graphene plasmon dispersion Characteristics} \label{GSP}
\setcounter{equation}{0}
\renewcommand{\theequation}{E{\arabic{equation}}}

\begin{figure}[htbp]
    \centering
%    \subfloat[]
    {\includegraphics[width=0.45\textwidth]{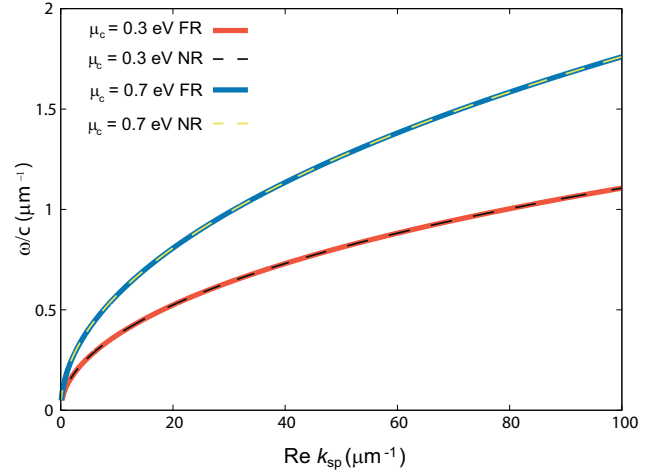}}
%    \hspace{1mm}
%    \subfloat[]{\includegraphics[width=0.45\textwidth]{OSA2.png}}
    \caption{GSP dispersion curves for $\mu_g=0.3,\,0.7$eV. Solid (dashed)  line is obtained by using the full retarded (FR) Eq. (\ref{sp}) (the non retarded (NR) Eq. (\ref{sp_cuasi})). }
    \label{alfasp}
\end{figure}

To obtain the dispersion characteristics of GSPs, we equate to zero  the denominator in Eq. (\ref{fresnelp}),
\begin{equation}\label{sp}
    \frac{\gamma^{(1)}}{\varepsilon_1}+\frac{\gamma^{(2)}}{\varepsilon_2}+\frac{Z_0 \sigma}{k_0} \frac{\gamma^{(1)}}{\varepsilon_1} \frac{\gamma^{(2)}}{\varepsilon_2}=0,
\end{equation}
and solve for $k_{||}$. 
In the non retarded regime, $k_0<<k_{sp}$, the wavenumber $k_{sp}$  can be approximated by \cite{jablan}
%Taking into account that $k_0<<k_{sp}$, we can apply the quasistatic approximation  neglecting  the term $k_0$ in $\gamma^{(j)}$ ($j=1,\,2$).
%
%Thus, the GSP propagation constant is written as, 
\begin{equation}\label{sp_cuasi}
    k_{sp}= i k_0\frac{\varepsilon_1 \varepsilon_2}{Z_0 \sigma (\varepsilon_1+\varepsilon_2)}= k_0\frac{\varepsilon_1 \varepsilon_2 \pi \hbar^2}{Z_0 e^2 \mu_g (\varepsilon_1+\varepsilon_2)} (\omega+i\gamma_g),
\end{equation}
where in the last equality we have used Eq. (\ref{intra2}). 
Figure \ref{alfasp} shows the dispersion curves obtained by solving Eq. (\ref{sp}) and the non retarded expression Eq. (\ref{sp_cuasi}) for various values of chemical potential,  $\mu_g=0.3,\,0.7$eV. We observe that results from Eq. (\ref{sp_cuasi}) agree well with the values obtained by solving the full retarded dispersion equation (\ref{sp}).  
\section{Numerical method} \label{Green_camino_de_integracion}

\begin{figure} [htbp!]
%\centering
\resizebox{0.6\textwidth}{!}
{\includegraphics{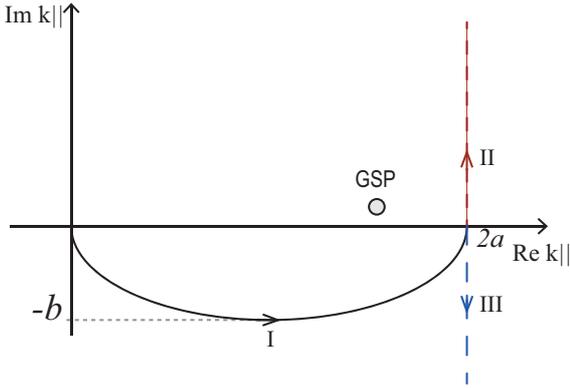}}
\caption{\label{fig:epsart} Singularities and path of integration in the complex plane $k_{||}=$Re $k_{||}+i $Im${k_{||}}$ for the electromagnetic fields. The original path, along the positive semi--axis, is deformed along an elliptical path (I) surrounding the singularities, together with the paths (II) and (III) parallel to the imaginary $k_{||}$ axis using Hankel functions. %An optical dipole emitter is inside a graphene--coated dielectric cylinder. The wire ($\varepsilon_1$, $\mu_1$ and surface conductivity $\sigma$) is embedded in a transparent medium with constitutive parameters $\varepsilon_2$, $\mu_2$.
}\label{integracion}
\end{figure}

The integration path in Eq. (\ref{G3p}) is set along the real and positive $k_{||}$ axis, so that the integral will be strongly  affected by singularities that are close to that axis. Pole singularities, zeroes of the denominator  in $r_p(k_{||} )$ coefficient, occur at  complex location %($k_{||}$ is a complex magnitude) 
and it represents  the propagation constant of the GSP mode. We  transform the original oscillatory integrand function into one to avoid the complex singularities which lie near the real $k_{||}$ axis and then we apply a numerical quadrature to calculate the field integrals. 
To do this, we  surround the pole  singularity by deforming the integration path into the complex plane as shown in Figure \ref{integracion}. The path I is an elliptical path starting at $k_{||}=0$ with the major semi--axis $k_{||}=a$ and the minor semi--axis $k_{||}=b$. In the region between  path I and the real axis the integrand function is analytical, thus the Cauchy's integral theorem implies that an integration on path I will be equal to the integral on the real axis from 0 to $2a$. The $a$ value should be chosen  large enough to surround  the pole singularities, thus $2a>k_{sp}$ must be fulfilled. %Unlike the metallic interface, the propagation constant of GSP can reach values up to 10 times greater than that corresponding to a photon  of the same frequency. 
%Therefore, 
In a first step %it is necessary  to
we divide the integration interval  into several subintervals and then  %to 
apply the numerical quadrature in each subinterval. We have implemented a 32 point Gauss Legendre quadrature in each of such sub--interval.  

Taking into account that,
\begin{equation}
J_n(z)=\frac{1}{2}[H_n^{(1)}(z)+H_n^{(2)}(z)],
\end{equation}
and the fact that the Hankel funtions of the first kind  $H_n^{(1)}(z)$ and the second kind $H_n^{(2)}(z)$  decrease faster as long as $|$Im $ z|$ increases in the sector Im $ z>0$ and Im $ z<0$, respectively, the remaining integration is carried out by deflecting the integration path from the real axis to a path parallel to the imaginary $k_{||}$ axis as shown in Figure \ref{integracion},  with Im $k_{||}>0$ for $H_n^{(1)}(k_{||}\rho)$ (path II) and with  Im $ k_{||}<0$ for $H_n^{(2)}(k_{||}\rho)$ (path III).  In the region between path II and the real axis the integrand has no pole singularities, thus Cauchy's integral theorem implies that the integral on a closed path in this region   will be zero.Therefore, the integral over path II in the direction shown in Figure \ref{integracion}  is equal  to that from $2a$ to$+\infty$ over the real axis. In a similar way, one can demonstrate that the integral over path III in the direction shown in Figure \ref{integracion}  is equal  to that from $2a$ to$+\infty$ over the real axis. In our implementation, we have used a   32 point Gauss Legendre quadrature  to calculate the field integrals on paths I, II and III.  

\end{document}